\documentclass[11pt,a4paper,twocolumn]{book}

\usepackage{jnpcs}
\usepackage{graphics} 
\usepackage{epsfig}

\ntitle{Convection in binary fluid mixtures: A model of four coupled amplitudes}

\nauthor{B.~Huke and M.~L\"ucke}

\naddress{\it Institute of Theoretical Physics, Saarland University,
Postfach   15 11 50, 66041   Saarbr\"ucken, Germany \\
{\small \rm
   E-mail:     huke@lusi.uni-sb.de }
}

\ndata{ }

\nabstract{ We consider the bifurcation scenario that is found 
in Rayleigh--B{\'e}nard convection of binary fluid mixtures like 
ethanol--water at positive 
separation ratios and small Lewis numbers leading to a bifurcation sequence
of square, oscillatory crossroll, stationary crossroll, and 
roll patterns. We propose a system of four coupled amplitudes that 
is capable to model all important properties. }

\nkeywords{ Rayleigh--B{\'e}nard convection, Soret effect, 
pattern selection, amplitude equations }

\nPACSnumbers{ 47.20.Bp; 47.20.Ky; 47.54.-r }
\begin{document}
\DeclareGraphicsExtensions{.jpg,.pdf,.mps,.png} 
\firstpage{1} \nlpage{2} \nvolume{4} \nnumber{1} \nyear{2001}
\def\nfpage{\thepage}
\thispagestyle{myheadings} \npcstitle

\section{Introduction}

The Rayleigh--B{\'e}nard system \cite{Ben00,Ray16} is one of the most popular 
model systems to study spontaneous structure formation emerging via 
instabilities. The system consists of a fluid layer confined between two 
parallel plates perpendicular to the direction of gravity, and exposed to a 
vertical temperature gradient that is established by keeping the upper and 
lower plate at two different temperatures $T_0 - \Delta T/2$ and 
$T_0 + \Delta T/2$ respectively.

Rayleigh--B{\'e}nard convection in one--component fluids at small and moderate 
temperature differences is well understood, mainly thanks to the work of 
Busse and co--workers in the 70s and 80s \cite{B67,CB74,CB78,BBC85}.
The bifurcation behavior becomes more complex in binary mixtures. 
Here, the concentration enters as an additional 
relevant dynamic field. Concentration differences are created in the presence 
of a non--vanishing Soret effect, a nonequilibrium effect that describes the 
driving of concentration currents by temperature differences.

The concentration couples back into the system's dynamics via variations that
determine the buoyancy force density.
When the lighter component of the mixture is driven into the direction of 
higher temperature, thereby enhancing the density gradient and further 
destabilizing the layer one speaks of a positive Soret effect, and of a 
negative Soret effect in the opposite case.

Convection in binary mixtures shows a very rich bifurcation behavior. For 
negative separation ratios convection rolls bifurcate backwards and 
time--dependent patterns, namely traveling waves and standing waves appear. 
The bistability of convective patterns and the ground state leads to the 
existence of localized structures and fronts.

In the case of positive separation ratios on which we will focus in this paper 
the two--dimensional roll structures are often replaced by squares at onset, but 
become the preferred pattern at higher temperature differences. For intermediate 
$\Delta T$ two types of crossroll structures can exist, a stationary 
and an oscillatory type that transfer stability from squares to rolls.    

There are few--mode models that satisfactorily explain the 
bifurcation properties for negative Soret effects \cite{HLM98}. 
However, for 
positive Soret effects a model with only a few degrees of freedom, that is
able to properly describe rolls, squares and crossrolls seems to be lacking.

In this paper we will present a system of four equations that describe all 
important properties of the square -- crossroll -- roll transition 
for the case of a positive Soret effect. 

The paper is organized as follows: after this introduction we discuss in 
Sec.~2 the Rayleigh--B{\'e}nard system and the bifurcation scenario for 
positive Soret effects in more detail. In Sec.~3, the main part, we  
present and discuss a model system that shows a similar behavior but 
incorporates an additional, artificial symmetry that makes it easier to identify 
the stationary solutions but leads to some qualitative disagreements with the 
full system. In Sec.~4 we add some terms that break this symmetry 
and demonstrate that then these differences disappear. We  
summarize our findings in Sec.~5 and provide an outlook.

\section{Bifurcation scenario}

The basic equations that couple convection velocity ${\bf u}$, temperature
$\theta$, concentration $c$ and pressure $P$ read in a dimensionless form 
\cite{PL84}
\begin{eqnarray}
\left( \partial_t + {\bf u} \cdot \nabla \right) {\bf u} &=& - \nabla P + 
\sigma \left[ \left( \theta + c \right) {\bf e}_z + \nabla^2 {\bf u} \right] 
\nonumber \\
\left( \partial_t + {\bf u} \cdot \nabla \right) \theta &=& R u_z + 
\nabla^2 \theta \\
\left( \partial_t + {\bf u} \cdot \nabla \right) c &=& R \psi u_z + 
L \left( \nabla^2 c - \psi \nabla^2 \theta \right)\nonumber \\
\nabla \cdot {\bf u} &=& 0 \;\; . \nonumber
\end{eqnarray}
The fields are given as deviations from the quiescent conductive ground state.
The direction of gravity is the negative $z$ direction. See ref. \cite{PL84} 
for details. There appear four 
different parameters in these equations: the Rayleigh number $R$ is the 
control parameter, the dimensionless temperature difference between upper 
and lower plate. In pure fluids, the critical Rayleigh number of the onset 
of convection is $R_c(\psi=0) = 1707.76$. One defines a reduced Rayleigh 
number $r = R/R_c(\psi=0)$. The Prandtl number $\sigma$ is the ratio of 
the timescales of heat and momentum diffusion. The Lewis number $L$ on the 
other hand is the ratio of the heat and concentration timescales. The 
separation ratio $\psi$ finally measures the strength of the Soret effect. 
In the conductive ground state it determines the ratio of the density 
gradients due to concentration and temperature.

In typical liquids it is $\sigma \approx 10$. Concentration diffusion is 
slow in ordinary liquid mixtures; a typical value for $L$, say, for 
ethanol--water is 0.01. $\psi$ varies with the mean concentration and mean 
temperature and can be both positive and negative in ethanol--water.

For positive $\psi$ three different bifurcation scenarios exist at small 
and moderate $r$. ($i$)~For very small $\psi$ and relatively large $L$ the Soret
generated concentration variations are small and easily diffused away. 
The mixture then still behaves qualitatively like a pure fluid. The first 
convection structure to be observed above onset takes the form of parallel, 
stationary rolls with alternating direction of rotation. Near onset, $u_z$, 
$\theta$, and $c$ can be described by a function $A \cos(k x) f(z)$ with $k$ 
being the wavenumber of the rolls and $f(z)$ being different for the three 
fields and defining the critical vertical profile. 

($ii$)~When $\psi$ is larger and/or $L$ is smaller the mixture begins to behave 
qualitatively different. The rolls are replaced by a square pattern near onset 
that can be written as a linear combination of equally strong rolls in $x$-- 
and $y$--direction, i.~e.,\  $\left[ A \cos(kx) + B \cos(ky) \right] f(z)$ with 
$A=B$. At higher $r$ however the advective mixing is strong enough to 
equilibrate the concentration field outside of some small boundary layers at the 
plates and between the rolls. The mixture again behaves like a pure fluid and 
squares lose their stability to rolls. The stability is transferred via 
a crossroll branch, a structure that, like squares, can be understood as a 
superposition of two perpendicular sets of convection rolls. In the case of 
crossrolls, however, the two amplitudes $A$ and $B$ are not equal and the
crossroll structure therefore lacks the square $x \leftrightarrow y$ symmetry. 
Like squares and rolls, these structures are stationary.  

\begin{figure}
\leavevmode
\centering
\includegraphics[width=8cm]{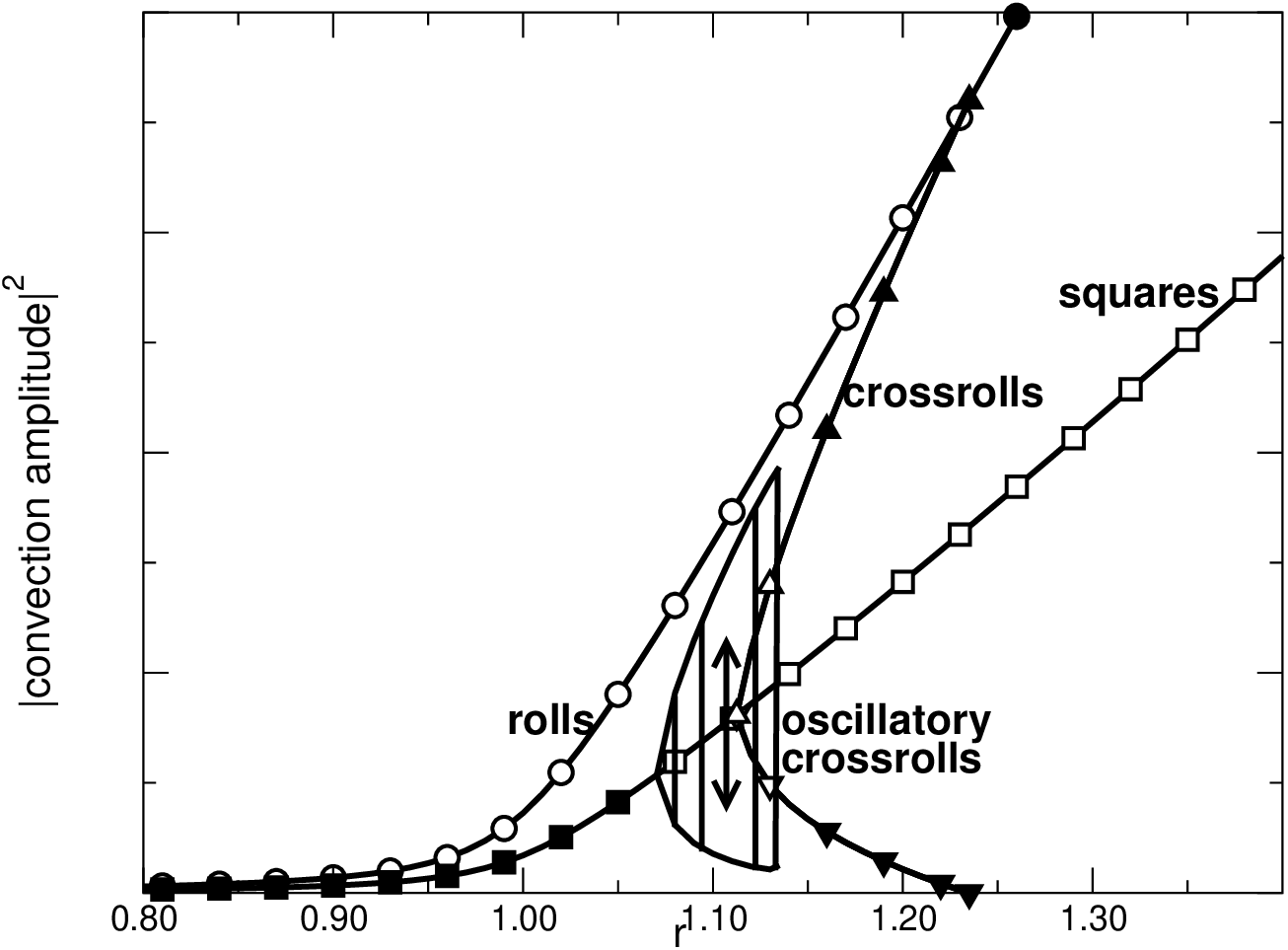}
\caption{Bifurcation scenario of the full hydrodynamic field equations for
 $\sigma=27$, $L=0.0045$, $\psi = 0.23$. The wave 
 number of the structures is $k = \pi$.}
\label{RBbif}
\end{figure}

($iii$)~At even smaller $L$, but easily realizable with liquid mixtures the 
scenario is again different as shown in Fig.~\ref{RBbif}. There the squared 
amplitude of the leading velocity mode of the $x$--roll set is plotted 
versus the reduced Rayleigh number $r$. Solid (open) symbols denote stable 
(unstable) structures. Squares (square symbols) are again stable at small $r$
but lose their stability against another type of crossrolls, an oscillatory 
one (lines without symbols). The bifurcation point where the stationary 
crossrolls (triangles) emerge still exists but at higher $r$, where the 
oscillatory instability has already taken place. In the oscillatory crossroll 
branch the amplitude of the $x$-- and $y$--roll set varies periodically in 
counterphase to each other such that when at one time the $x$--rolls are 
stronger, half a period later the $y$--rolls will be the stronger set. The 
two branches  in Fig.~\ref{RBbif} show 
the maximal and minimal amplitudes of these time dependent structures.
The two stationary crossroll branches represent the two different types: In the
upper branch the $x$--roll component is the dominant one and their amplitude 
grows whereas the amplitude of the $y$--roll component gets smaller. When
the latter component vanishes the crossroll ends at a bifurcation point on 
the $x$--roll branch (circles). The lower branch belongs to the type of 
crossrolls where the $y$--roll component is the dominant one. Now the plotted 
$x$--roll component gets smaller until the structure ends on the $y$--roll 
branch, i.~e., on the axis in Fig.~\ref{RBbif} since there only the strength of
the $x$--rolls is plotted. The stationary crossrolls gain stability when the 
oscillatory crossrolls vanish and transfer the stability to the rolls.
The existence of all these structures have been demonstrated in experiments
\cite{GPC85,MS86,BCC90}.

\begin{figure}
\leavevmode
\centering
\includegraphics[width=8cm]{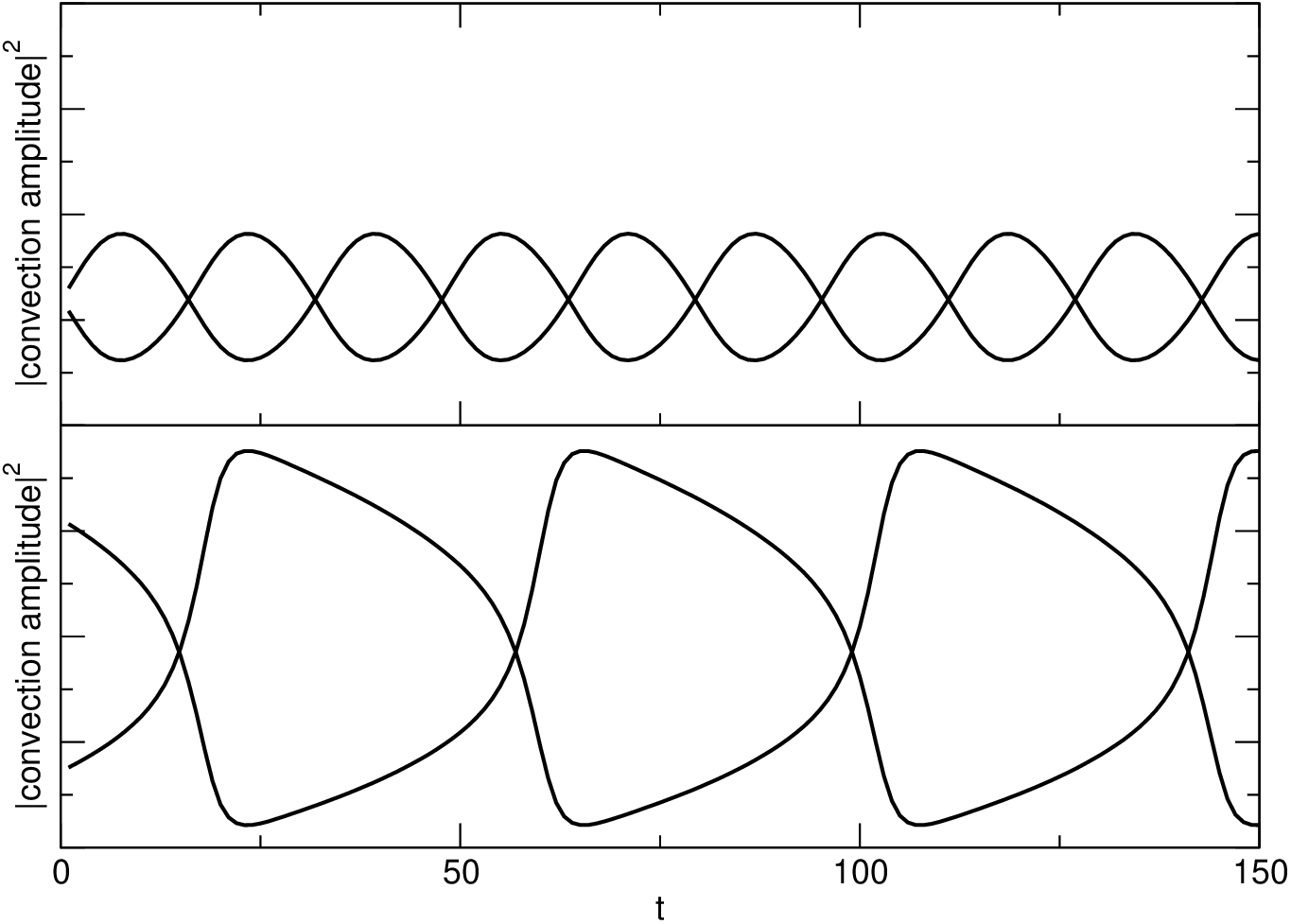}
\caption{Oscillatory crossrolls, squared amplitude of the $x$-- and $y$--roll
component for the parameters of Fig.~\ref{RBbif} and $r=1.08$ (top), $r=1.13$
(bottom).}
\label{oscifull}
\end{figure}

The time--dependence of $A$ and $B$ is plotted for two oscillatory crossrolls
in Fig.~\ref{oscifull}. It shows the main characteristics of these structures
near the beginning and the end of their $r$--range of existence.
While the oscillation is very harmonic at smaller $r$ it becomes very
anharmonic at higher $r$. The frequency decreases. Finally, the oscillatory 
crossrolls end in an entrainment process in a very narrow $r$--interval 
\cite{JHL98}.

The crossroll structures exist around $r = 1$, i.~e.\ the critical Rayleigh number 
of the pure fluid. In the same region the branches of rolls and squares bend 
upwards. Before and after this transition interval the curves are almost
linear but with very different slopes. These two regimes at small and large $r$
are called the Soret and the Rayleigh regime respectively. In the Soret region
the convection can be well explained by an amplitude equation model including 
only cubic terms. In the Rayleigh region on the other hand the concentration 
field has been equilibrated and the convection amplitudes agree well with those 
of the pure fluid. In the Rayleigh region slightly above $r=1$ the pure fluid 
still obeys its own amplitude equation and so does therefore indirectly also the 
mixture \cite{HLBJ00}. We will make use of the fact that the bifurcation behavior 
of the mixture can in fact be described by two different amplitude equations in 
two different $r$--intervals in the next section.   

\section{The simplified model}

In this section we want to present a first attempt to explain the bifurcation
scenario as described above by means of a simple model. The full picture
cannot be captured by simple amplitude equations derived in an expansion
around the onset in combination with a multi--scale 
analysis from the basic equations. Although a sequence of stable squares, 
forward bifurcating stable (stationary) crossrolls, and finally stable rolls 
may be described with a system of two coupled quintic amplitude equations 
\cite{CK92}, the oscillatory crossrolls are not contained in such a model. When 
$A$ is the real amplitude of $x$--rolls, referring to a critical mode 
$A \cos( k x) f(z)$, and $B$ is the corresponding mode in $y$--direction 
then a system of two coupled amplitude equations must read for reasons of 
symmetry

\begin{eqnarray}
d_t A &=& \mu A - F \left( A^2, B^2 \right) A \nonumber \;\; , \\
d_t B &=& \mu B - F \left( B^2, A^2 \right) B \label{generalAE} \;\; ,
\end{eqnarray}
with some arbitrary function $F$. $F$ is real since the critical 
perturbations lead to stationary patterns. For squares, it is $A = B$ and 
the Jacobian of the right hand sides of the above equations will be symmetric.
Therefore, complex eigenvalues that would mark the appearance of oscillatory 
crossrolls are not possible along this branch.  

In order to reflect the existence of the Soret and Rayleigh regime we construct
a model of two coupled sets of amplitude equations. We define two different 
amplitudes $A_1$ and 
$A_2$, both describing rolls in $x$--direction of different nature: solutal 
rolls that exist in the Soret region due to the destabilizing effect of the 
initial concentration gradient and which are described by a mode $A_1 \cos (k x) f(z)$ 
and the thermal rolls $A_2 \cos (k x) g(z)$ that exist in the Rayleigh 
regime environment of almost homogeneous concentration and which are driven by
the thermal stress alone. $B_1$ and $B_2$ are the corresponding amplitudes 
for the rolls in $y$--direction.

The most simple model would consist of two sets of equations as in 
(\ref{generalAE}), including only cubic terms to couple the $A$-- and 
$B$--rolls and additional cubic terms to couple the index 1 and index 2 
structures:
\begin{eqnarray}
d_t A_1 &=& \mu_1 A_1\nonumber \\ 
\nonumber
&& \!\!\!\! - A_1 ( b_1 A_1^2 + c_1 B_1^2 + d_1 A_2^2 + e_1 B_2^2 ) \;\; , \\
d_t A_2 &=& \mu_2 A_2 \nonumber\\ 
\label{fullequationssimple}
&& \!\!\!\! - A_2 ( b_2 A_2^2 + c_2 B_2^2 + d_2 A_1^2 + e_2 B_1^2 ) \;\; , \\ 
d_t B_1 &=& \mu_1 B_1 \nonumber \\ 
\nonumber
&& \!\!\!\! - B_1 ( b_1 B_1^2 + c_1 A_1^2 + d_1 B_2^2 + e_1 A_2^2 ) \;\; , \\
d_t B_2 &=& \mu_2 B_2 \nonumber\\ 
\nonumber
&& \!\!\!\! - B_2 ( b_2 B_2^2 + c_2 A_2^2 + d_2 B_1^2 + e_2 A_1^2 ) \;\; .
\end{eqnarray}
These equations do not define an amplitude equation model in a strict sense
since they are not meant to result from a controlled expansion of the dynamics 
near the bifurcation threshold of the full system. Instead, in order to describe 
the actual convection behavior we have two bifurcation points here, as it is
displayed by the two different control parameters $\mu_1$ and $\mu_2$.
These parameters are not independent however. Since the only control 
parameter in the Rayleigh--B{\'e}nard system is $r$, the parameters $\mu_i$ 
must be functions of each other. We set
\begin{equation}
\mu_1 = \mu \;\; , \qquad \qquad \mu_2 = a \mu - \mu_0 \;\; .
\end{equation} 
The two bifurcation points are at
$\mu_1 = 0$ and at $\mu_2 = 0$. Since for $L \ll 1$ solutal rolls grow on a 
much slower timescale than thermal rolls one has to have $a \gg 1$.
When $\mu = -1$ refers to $r = 0$ it has to be $\mu_2 = 0$ at $\mu \gg 1$ 
since the onset of convection in mixtures lies at much smaller $r$ than 
the onset $r=1$ in a pure fluid. For the plots we set $a = 10$ and
$\mu_0 = 100$. That means that the timescale of the solutal rolls is assumed 10 
times larger than the one of the thermal rolls. The latter appear only
at $\mu > 10$, or 11 times the critical Rayleigh number of the solutal rolls.
The ratios found in the numerical investigation of the full system are 
typically even larger for realistic parameter combinations.     

We assume all parameters $b_1, b_2, ... e_2$ to be positive. A
different sign of any of these would lead to unwanted features in the
bifurcation scenario. Since $a \gg 1$ we can assume that for any two 
parameters $p$, $q$ $\in \{b_1, b_2, ... e_2\}$ the inequality $p < a q$ 
holds. This assumption will simplify our discussion below. We can set 
$b_1 = b_2 = 1$ by normalizing the amplitudes accordingly.

The equations~(\ref{fullequationssimple}) show several symmetry properties. 
They are invariant under an exchange of $A$ and $B$ amplitudes, i.e.,
under exchange of $x$ and $y$. The absence of quadratic terms guarantees the
preservation of the upflow--downflow symmetry of the convective fields. A 
change of the sign of only $B_1$ and $B_2$ or only $A_1$ and 
$A_2$ corresponds to a shift of the $x$-- or $y$--rolls by half a 
wavelength. 

Furthermore, it is even possible to change the sign of only one amplitude. 
However, this symmetry is artificial. The real convection patterns must be considered 
as a nonlinear superposition of the solutal and thermal patterns. Changing, 
e.~g., in $x$--rolls the direction of rotation of the solutal contribution
while leaving the thermal one unchanged is certainly not a symmetry operation. 
Nevertheless we will add terms that break this symmetry 
of eqs.~(\ref{fullequationssimple}) only in the next section.

Although the bifurcation scenario of the simplified model (\ref{fullequationssimple})
disagrees qualitatively with that of the full hydrodynamic field equations in
some points it nevertheless facilitates to understand some other features. 
Furthermore, the simplified model easily 
allows to get analytical solutions for the stationary structures: Every single
mode can be set independently to zero. The remaining equations can in the
stationary case then be rewritten as a linear system with the squares of
the nonzero amplitudes as variables. The unique solution is a linear function 
of $\mu$, and physically relevant in a $\mu$--interval where all amplitude
squares are $\geq 0$. We can conclude that there are at maximum 16 different
stationary solutions of eqs.~(\ref{fullequationssimple}) because there are 
$2^4$ possible ways to set some amplitudes to zero.

\subsection{Thermal and solutal rolls}

\begin{figure}[h!]
\leavevmode
\centering
\includegraphics[width=8cm]{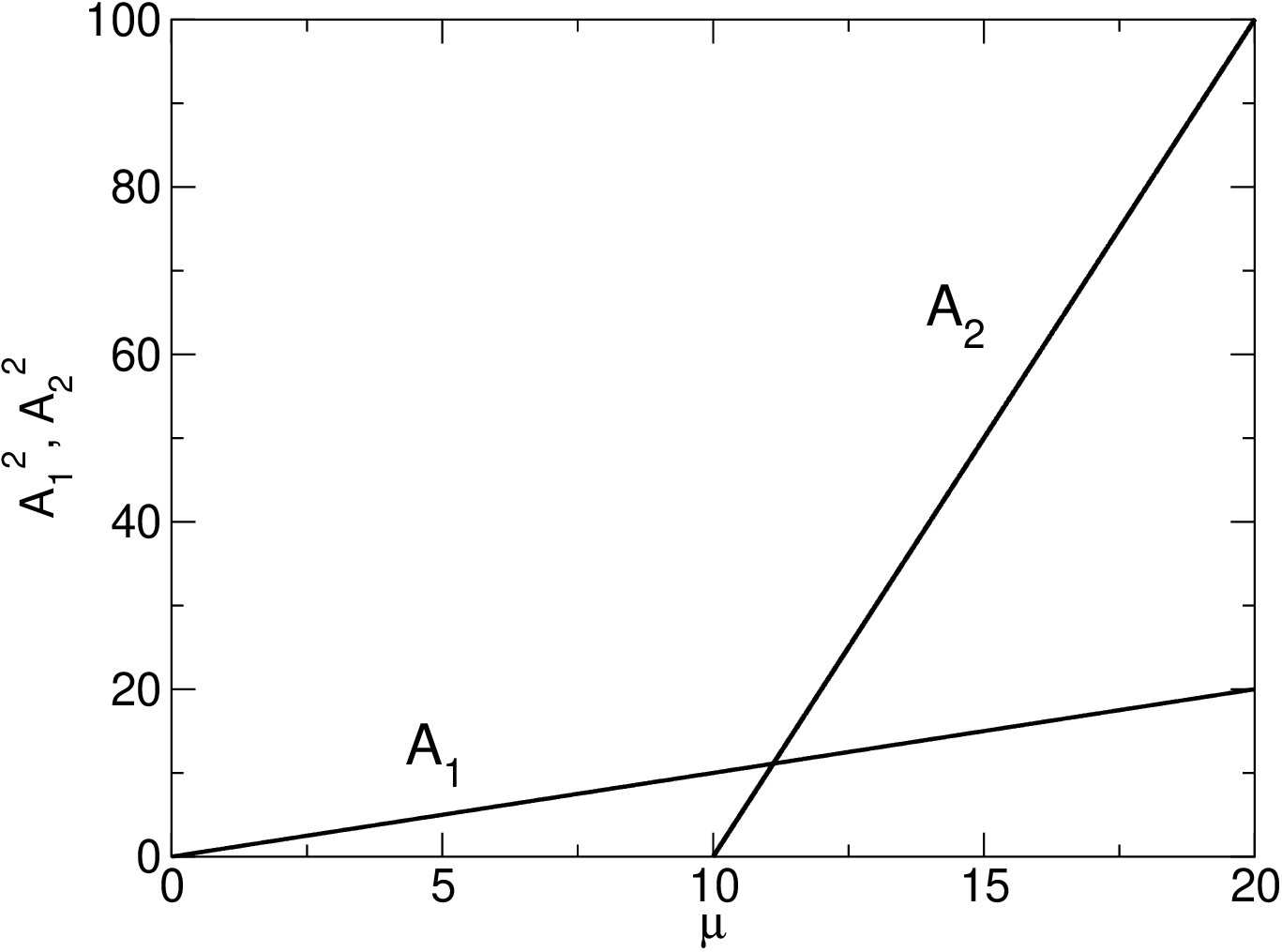}
\caption{Bifurcation of pure solutal $(A_1)$ and pure thermal $(A_2)$ rolls of the
model eqs.~(\ref{fullequationssimple}). Parameters are $a=10$, $\mu_0 = 100$,
and $b_1 = b_2 = 1$.}
\label{tworollfig}
\end{figure}

The first kind of structures we discuss are $x$--rolls with 
$B_1 = B_2 = 0$ as solutions of the subsystem
\begin{eqnarray}
d_t A_1 &=& \mu_1 A_1 - A_1 \left( b_1 A_1^2 + d_1 A_2^2 \right) \;\; ,
\nonumber \\
d_t A_2 &=& \mu_2 A_2 - A_2 \left( b_2 A_2^2 + d_2 A_1^2 \right) \;\; .
\label{eqrollsystem}
\end{eqnarray}
Since $b_1, b_2 > 0$, both pure solutal or $A_1$--rolls 
\begin{equation}
A_1^2 = \frac{\mu_1}{b_1}\;\; , \;\; A_2 = 0 \;\;\;\; (\mu_1 = \mu > 0)\;\; , 
\end{equation}
and pure thermal or $A_2$--rolls
\begin{equation}
A_2^2 = \frac{\mu_2}{b_2}\;\; , \;\; A_1 = 0 \;\;\;\; 
(\mu_2 = a \mu - \mu_0 > 0) \;\; ,
\end{equation}
exist as forward bifurcating solutions. They are plotted in 
Fig.~\ref{tworollfig} for $b_1 = b_2 = 1$ and $a$, $\mu_0$ as above.

As the only convective solution for small $\mu$, the solutal rolls are stable 
at onset. But the linear stability analysis shows that they lose stability in 
favor of the stronger thermal rolls at larger $\mu$. The stability is 
transfered via a third kind of rolls, the intermediate or $A_1A_2$--rolls
given by
\begin{eqnarray}
A_1^2 &=& \frac{b_2 \mu_1 - d_1 \mu_2}{b_1 b_2 - d_1 d_2} \nonumber \\
A_2^2 &=& \frac{b_1 \mu_2 - d_2 \mu_1}{b_1 b_2 - d_1 d_2} \;\; .
\label{interrolls}
\end{eqnarray}
These structures exist if there is an $\mu$--interval where both right--hand 
sides are $> 0$, what is always the case. If $b_1 b_2 - d_1 d_2 > 0$ they 
bifurcate forward from solutal to thermal rolls.

\begin{figure}[h!]
\leavevmode
\centering
\includegraphics[width=8cm]{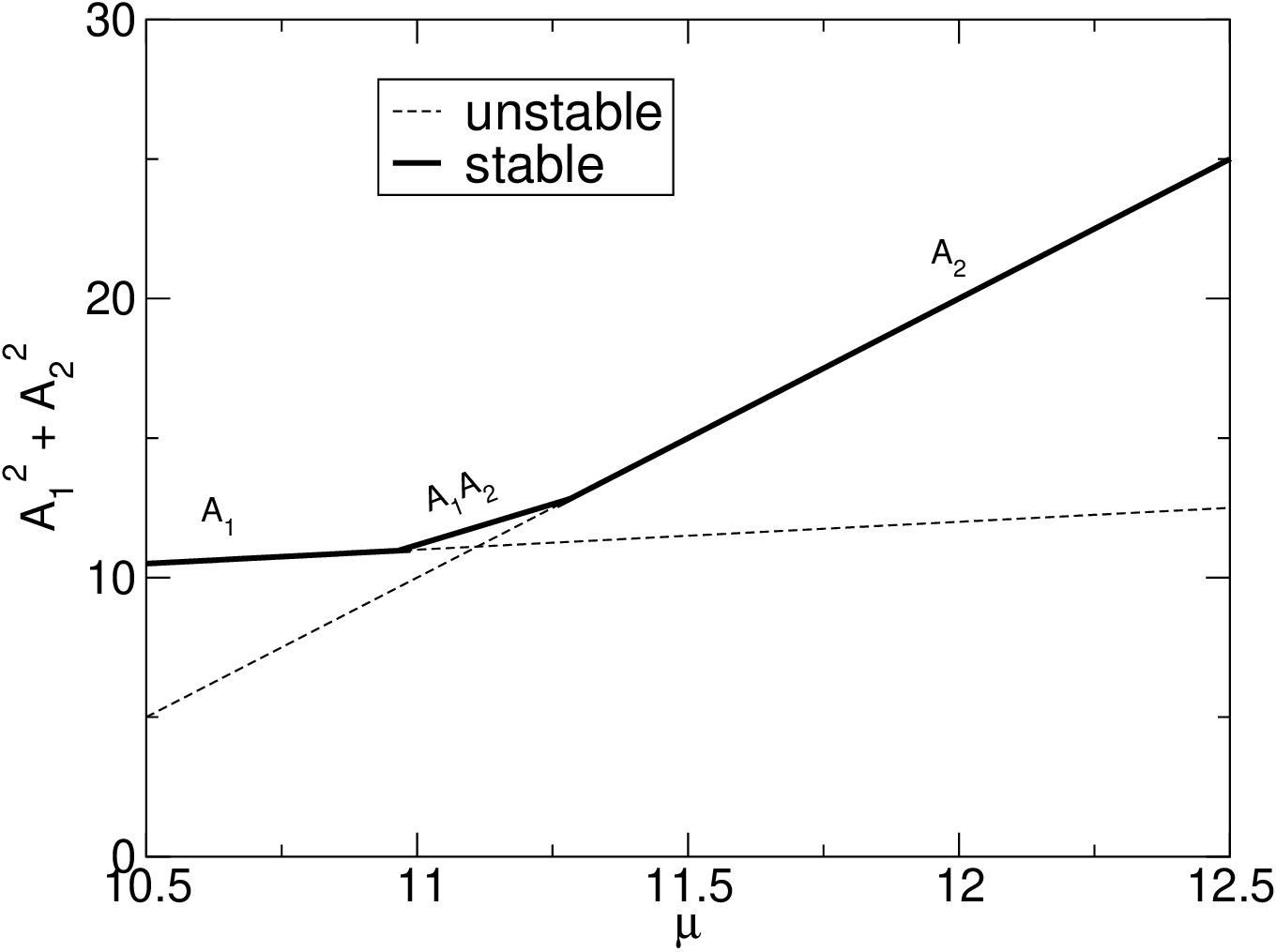}
\caption{The three types of stationary rolls solutions of the model 
eqs.~(\ref{fullequationssimple}):
 Solutal rolls ($A_1$), 
thermal rolls ($A_2$) and intermediate rolls ($A_1A_2$) plotted as
$A_1^2+A_2^2$ versus $\mu$. Solid (dashed) lines 
denote stable (unstable) structures. Parameters are $d_1=d_2=0.88$ and 
otherwise as above.}
\label{threerollfig}
\end{figure}

A bifurcation diagram showing all three kinds of solutions and also their
stability properties within the restricted system (\ref{eqrollsystem})
is shown in Fig.~\ref{threerollfig}. Plotted is only the interesting 
$\mu$--range where 
stability is transferred. The stable branches already look like a caricature of
the roll branch in the full system with different slopes in the Soret and 
Rayleigh regions. However, there are also important qualitative differences. 
In the full system, the roll branch consists of one continuous structure, 
without the two bifurcation points that are found in the model system. These 
bifurcation points will disappear when we break the artificial symmetry present 
in the model system.

\subsection{Thermal and solutal squares}

\begin{figure}[h!]
\leavevmode
\centering
\includegraphics[width=8cm]{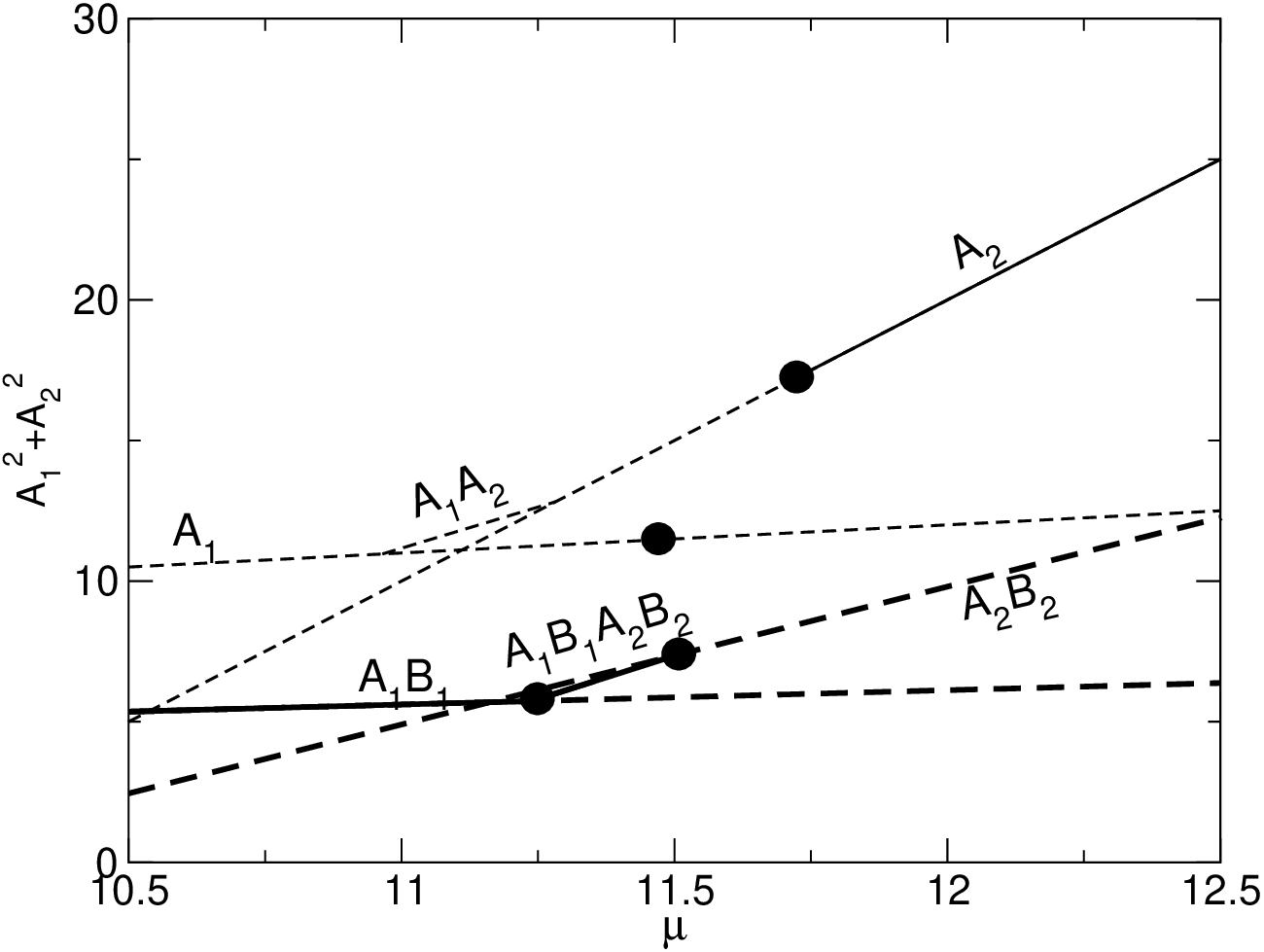}
\caption{Square and roll structures of the model 
eqs.~(\ref{fullequationssimple}). Squares (rolls) are represented by thick 
(thin) lines. Solid (dashed) lines  denote stable (unstable) structures.
Bifurcation points of further interest are marked by circles. Parameters are
$c_1 = 0.96$, $c_2 = 1.04$, $e_1 = 0.68$, $e_2 = 1.28$ and otherwise as above.}
\label{RQplot}
\end{figure}

A solution has square symmetry when $A_1 = B_1$ and $A_2 = B_2$. Analogous to
rolls there are three different types of squares, namely solutal or 
$A_1B_1$--squares
\begin{equation}
A_1^2 = B_1^2 = \frac{\mu_1}{b_1 + c_1}\;\; , \;\; A_2 = B_2 = 0 \;\; , 
\end{equation}
thermal or $A_2 B_2$--squares
\begin{equation}
A_2^2 = B_2^2 = \frac{\mu_2}{b_2 + c_2}\;\; , \;\; A_1 = B_1 = 0 \;\; ,
\end{equation}
and intermediate $A_1 B_1 A_2 B_2$--squares
\begin{eqnarray}
A_1^2 = B_1^2 &=& 
\frac{(b_2 + c_2)\mu_1 - (d_1 + e_1)\mu_2}{\Delta} \;\; , \nonumber \\
A_2^2 = B_2^2 &=& 
\frac{(b_1 + c_1)\mu_2 - (d_2 + e_2)\mu_1}{\Delta} \;\; , 
\end{eqnarray}
where 
\begin{equation}
\Delta = (b_1+c_1) (b_2+c_2) - (d_1+e_1) (d_2+e_2)\;\; .
\end{equation}
Under the general requirements stated above the intermediate squares
exist and they bifurcate forward from solutal to thermal squares if 
$\Delta >0$. The bifurcation branches of the different square and roll 
structures are plotted in Fig.~\ref{RQplot}. 

The stability properties of the different types of squares against each
other are analogous to those of the rolls. But these are not all types
of possible instabilities. Since our model has four modes four eigenvalues
have to be investigated for all structures found so far. We will briefly
discuss the respective results.

Some instabilities are connected to the bifurcation points that are known 
already.
For the squares only those instabilities that can break the 
($x \leftrightarrow y$)--symmetry are still of interest. It turns out that
solutal squares are stable against ($x \leftrightarrow y$)--antisymmetric 
$(\delta A_1,\delta B_1)$--perturbations everywhere if and only if $c_1 < b_1$. This 
inequality is satisfied in the example case and therefore squares are stable at 
onset as desired. 
Similarly, thermal squares are stable against a type of 
$(\delta A_2,\delta B_2)$--perturbations only if $c_2 < b_2$. This is not the case and 
consequently the thermal squares have been marked as unstable in 
Fig.~\ref{RQplot}.

These requirements have also consequences for the stability of rolls: Solutal 
rolls, although stable against other $x$--roll structures for small $\mu$  
are unstable against $\delta B_1$--perturbations if and only if $c_1 > b_1$ as 
required for solutal squares to be stable. On the other hand, thermal rolls 
are stable against $\delta B_2$--perturbation when $c_2 < b_2$ and thermal squares
are unstable. Therefore, only the thermal rolls are marked as stable in 
Fig.~\ref{RQplot}.

There are two other types of instability of rolls that have to be discussed, 
namely the instability of solutal $A_1$--rolls against $\delta B_2$--perturbations 
and vice versa, of thermal $A_2$--rolls against $\delta B_1$--perturbations. Both
instabilities generate new bifurcation points that have been marked by 
large circles in the figure. Their position depends on the $d_i$ and $e_i$. 
Due to the fact that $e_2 > d_2$, the new bifurcation point on the solutal roll 
branch lies at higher $\mu$ than the bifurcation point to the intermediate 
rolls, where the solutal rolls already lose stability. Since $e_1 < d_1$, on the
other hand, the new bifurcation point for thermal rolls lies on the otherwise 
stable part of the thermal roll branch, destabilizing it towards the region 
of smaller $\mu$ already before the bifurcation point towards the 
intermediate rolls. 

Placing the new bifurcation points this way no new instabilities are to be 
expected for the intermediate rolls. They inherit stability against 
$\delta B_2$--perturbations and instability against $\delta B_1$--perturbations
 from the 
solutal rolls and transfer these properties to the thermal rolls. 

It should be pointed out that by choosing the parameters accordingly it is
also possible to stabilize the rolls in the whole $\mu$--interval against
perturbations in $y$--direction, which is the scenario that can be found at
large $L$ and small $\psi$ where the square patterns are always unstable.

Concerning the square structures we have to remark that at the bifurcation
point where the intermediate squares branch off from the solutal squares
in fact not one but two eigenvalues go through zero for symmetry reasons,
namely both $\delta A_2$-- and $\delta B_2$--perturbations. Two types of
 crossroll 
branches appear here plus the intermediate squares as nonlinear 
combination of the two. The same happens at the bifurcation point of the 
intermediate and thermal squares. Here also two new patterns emerge, 
connected to growing $\delta A_1$-- and $\delta B_1$--perturbations. These 
bifurcation points are
also marked by circles as a reminder of the new branches that still have 
to be discussed.

\subsection{Crossroll structures}

We have already pointed out above that at maximum 16 different stationary 
structures exist within the model (\ref{fullequationssimple}). The most simple 
one is the ground state where all 
amplitudes vanish. Furthermore we have already discussed the $A_1$--, 
$A_2$--, and $A_1A_2$--rolls that have their counterpart in 
$y$--direction, i.~e.,  the $B_1$--, \mbox{$B_2$--,} and $B_1B_2$--rolls.
Together with the \mbox{$A_1B_1$--,} \mbox{$A_2B_2$--,} and 
$A_1B_1A_2B_2$--squares these
are 10 structures so far. The remaining six are crossroll structures,
i.~e., three--dimensional patterns lacking the 
($x \leftrightarrow y$)--symmetry of squares.

The first two of these structures are the $A_1B_2$--crossrolls and the 
$B_1A_2$--crossrolls that can be mapped onto each other via the 
($x \leftrightarrow y$)--symmetry operation. The $A_1B_2$--crossrolls
connect the $A_1$--rolls and the $B_2$-rolls at the newfound bifurcation
points for the roll structures. Note that $B$--roll branches would lie on the 
$\mu$--axis in the plots since only $A_1^2 + A_2^2$ is plotted. The stability 
analysis reveals two more bifurcation points on these branches as it has to be 
since they connect the stable thermal rolls to the twofold unstable solutal 
rolls.

The next pair are the $A_1 B_1 A_2$-- and $A_1 B_1 B_2$--crossrolls that
branch off at the connection between solutal and intermediate squares and
end up at one of the bifurcation point of the $B_1A_2$-- and 
$A_1 B_2$--crossrolls, respectively.

The $A_1 A_2 B_2$-- and $B_1 A_2 B_2$--crossrolls finally emerge at the
intermediate squares/thermal squares bifurcation point. These too end
up at the $A_1B_2$-- and $B_1 A_2$--crossroll branches, at the other two 
remaining bifurcation points.

\begin{figure}[h!]
\leavevmode
\centering
\includegraphics[width=8cm]{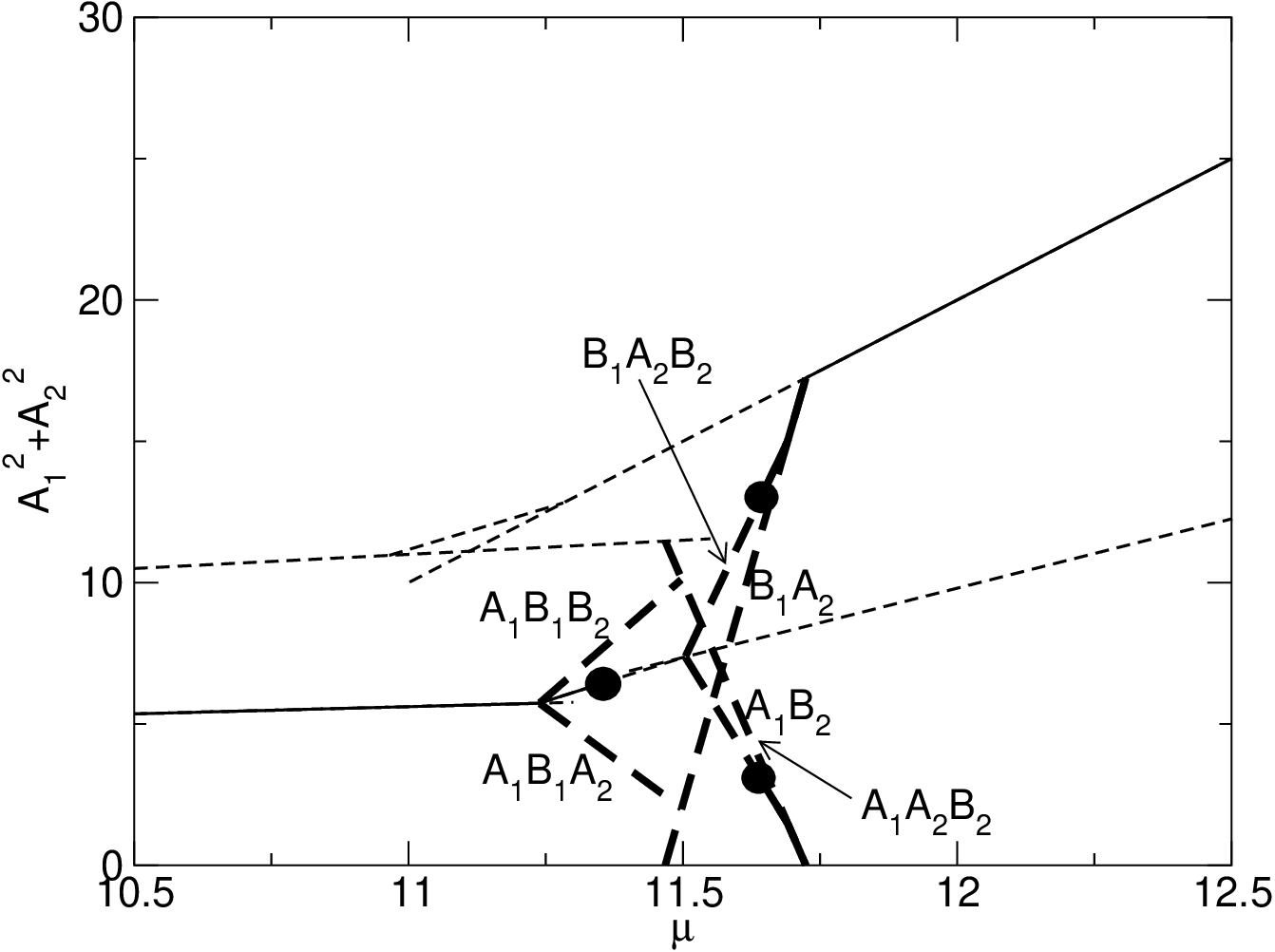}
\caption{Stationary crossroll solutions of eqs.~(\ref{fullequationssimple}). 
Thick (thin) lines represent
crossrolls (rolls and squares). Solid (dashed) lines denote stable (unstable) 
structures. Oscillatory instabilities occur at the circles.}
\label{allstationary}
\end{figure}

That completes the description of the stationary structures. The results
are shown in Fig.~\ref{allstationary}, together with the result of the 
full stability analysis. The unimportant part of the solutal 
and thermal roll and square branches have been cut off for better visibility. 
At small $\mu$ solutal squares are stable. They lose their stability where
two types of crossrolls ($A_1B_1A_2$, $A_1B_1B_2$) and the intermediate 
squares branch off. The intermediate squares gain stability whereas the 
crossrolls are unstable. Taking only stationary structures into account the 
intermediate squares remain stable until a second bifurcation point, where 
again two crossroll branches emerge ($A_1A_2B_2$, $B_1A_2B_2$) that are
now stable. The 
thermal squares on the other hand do not gain stability. Next, the stable 
crossrolls meet a third pair ($A_1B_2$, $B_1A_2$) that finally transfers
stability to the thermal rolls. The bifurcation diagram looks qualitatively
different for other parameter combinations but the inequalities that have to 
be fulfilled to produce a diagram as in Fig.~\ref{allstationary} can be 
derived easily.

The picture is still not complete however. Although all stationary 
instabilities have been identified now, there might be more bifurcation points
where time dependent patterns emerge. From the otherwise stable branches only
the mixed squares and the $B_1A_2B_2$-- and $A_1A_2B_2$-crossrolls can exhibit
oscillatory instabilities. For the given parameters such instabilities do 
indeed occur. They destabilize the parts of the branches
between the circles in Fig.~\ref{allstationary}. One can
easily guess that this is the region where oscillatory crossrolls can be 
found. But we want to discuss these structures in more detail only in the
full model.

\section{The full model}

The bifurcation diagram we found does already show some of the properties of
the scenario we want to explain but turned out to have unwanted 
bifurcation points. We will now show how including terms that break the 
artificial extra
symmetry leads to a much better model. There are four more cubic terms
preserving the symmetries $(A_1,A_2) \leftrightarrow (B_1,B_2)$, 
$(A_1, A_2) \leftrightarrow (-A_1, -A_2)$, and $(B_1, B_2) \leftrightarrow 
(-B_1, -B_2)$. When "..." denotes the old right hand sides of 
(\ref{fullequationssimple}) we enlarge the model as follows
\begin{eqnarray}
d_t A_1 &=& ... \nonumber\\
\nonumber
&& \!\!\!\!\!\! - A_2 ( \beta_1 A_1^2 + \gamma_1 B_1^2 + \delta_1 A_2^2 + 
\epsilon_1 B_2^2 ) \;\; , \\
d_t A_2 &=& ... \nonumber\\
\label{fullequations}
&& \!\!\!\!\!\! - A_1 ( \beta_2 A_2^2 + \gamma_2 B_2^2 + \delta_2 A_1^2 + 
\epsilon_2 B_1^2 ) \;\; , \\
d_t B_1 &=& ... \nonumber \\
\nonumber
&& \!\!\!\!\!\! - B_2 ( \beta_1 B_1^2 + \gamma_1 A_1^2 + \delta_1 B_2^2 + 
\epsilon_1 A_2^2 ) \;\; , \\
d_t B_2 &=& ... \nonumber\\
\nonumber
&& \!\!\!\!\!\! - B_1 ( \beta_2 B_2^2 + \gamma_2 A_2^2 + \delta_2 B_1^2 + 
\epsilon_2 A_1^2 ) \;\; .
\end{eqnarray}
The new equations do not allow a sign reversal of individual amplitudes 
anymore. However, the presence of the $\beta_i$, $\gamma_i$, and $\epsilon_i$
alone still allows pure solutal or thermal stationary roll solutions. This 
leads again to unwanted bifurcation points. On the other hand, when 
$\delta_i \neq 0$ such solutions become impossible. We will therefore discuss 
only these two and set the other six to zero. Thus the model we discuss here is
\begin{eqnarray}
d_t A_1 &=& \mu_1 A_1 - \delta_1 A_2^3 \nonumber \\ 
\nonumber
&& \!\!\!\! - A_1 ( b_1 A_1^2 + c_1 B_1^2 + d_1 A_2^2 + e_1 B_2^2 ) \;\; , \\
d_t A_2 &=& \mu_2 A_2 - \delta_2 A_1^3 \nonumber\\ 
\label{fulleqns2}
&& \!\!\!\! - A_2 ( b_2 A_2^2 + c_2 B_2^2 + d_2 A_1^2 + e_2 B_1^2 ) \;\; , \\ 
d_t B_1 &=& \mu_1 B_1 - \delta_1 B_2^3 \nonumber \\ 
\nonumber
&& \!\!\!\! - B_1 ( b_1 B_1^2 + c_1 A_1^2 + d_1 B_2^2 + e_1 A_2^2 ) \;\; , \\
d_t B_2 &=& \mu_2 B_2 - \delta_2 B_1^3 \nonumber\\ 
\nonumber
&& \!\!\!\! - B_2 ( b_2 B_2^2 + c_2 A_2^2 + d_2 B_1^2 + e_2 A_1^2 ) \;\; .
\end{eqnarray}

It is instructive to first take a look at the impact that the two new 
symmetry--breaking terms have on the eigenvalues of the important branches
as they emerge from the linear stability analysis. 
\begin{figure}[h!]
\leavevmode
\centering
\includegraphics[width=8cm]{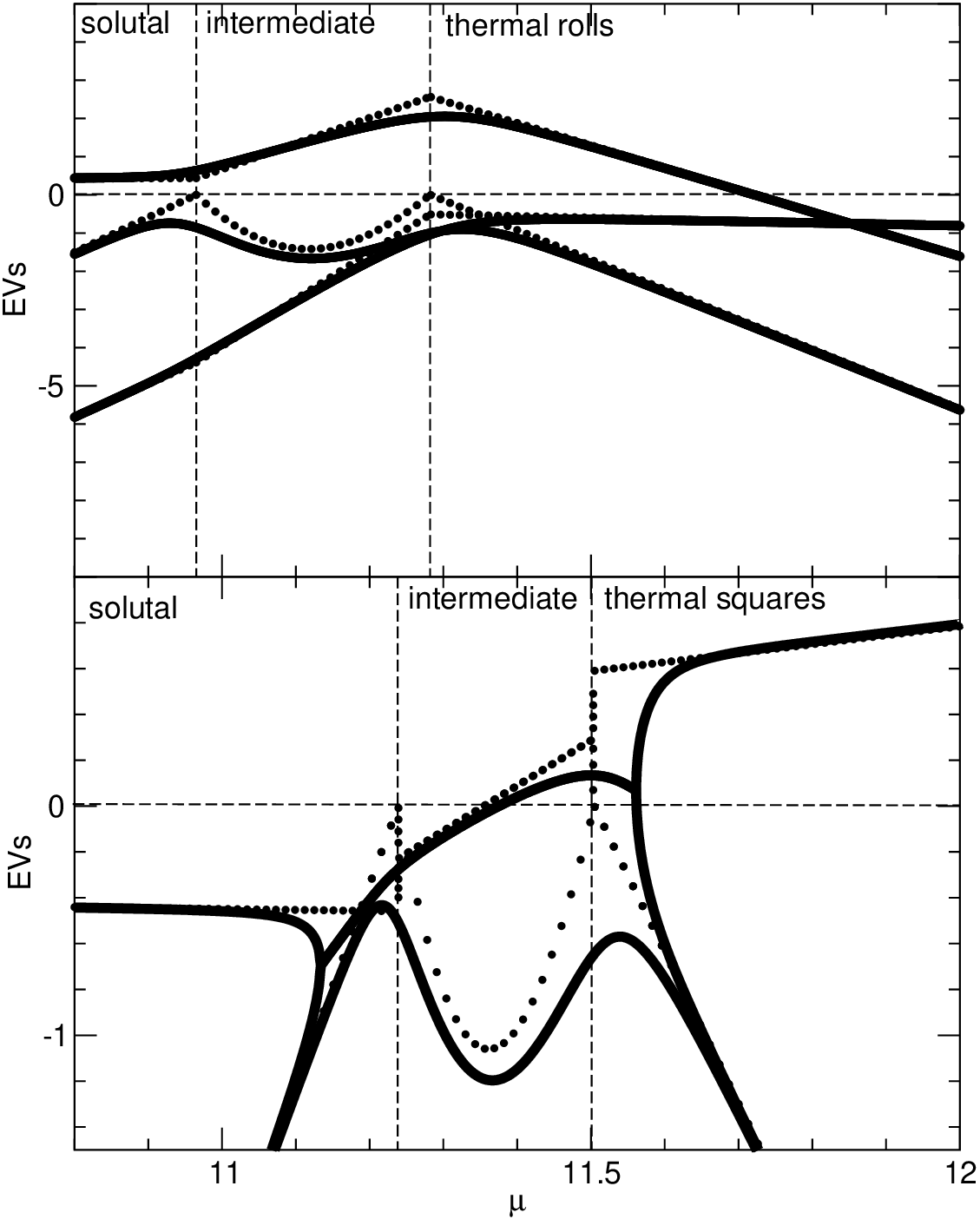}
\caption{Eigenvalues (EVs) along the important parts of the different roll and
square branches for $\delta_1 = \delta_2 = 0.01$. Dots denote the simplified 
system with $\delta_i=0$.}
\label{eigvals}
\end{figure}
Fig.~\ref{eigvals} shows the real parts of the three most important 
eigenvalues for both rolls and squares for small values 
$\delta_1 = \delta_2 = 0.01$ and $\delta_i = 0$ in comparison. Plotted are 
the intermediate structures in their $\mu$--interval of existence and the 
solutal and thermal structures below and above, respectively. In the upper 
plot, one can see how one eigenvalue of the solutal and thermal rolls becomes 
zero in the simplified system at the points where the intermediate rolls 
emerge. The referring curves would cross the zero axis here if the picture 
would not switch to the intermediate rolls. For $\delta_i \neq 0$ these 
bifurcations become imperfect. The curves become smooth and do not touch
the zero axis anymore such that the two bifurcation points vanish. The 
respective parts
of the solutal, intermediate, and thermal roll branches fuse into one single
branch. On the other hand, the bifurcation point to the crossrolls at 
$\mu \approx 11.7$ where the rolls finally become stable does not go away when 
$\delta_i \neq 0$.

In the lower plot one can observe a similar behavior for the square eigenvalues.
In the simplified system the solutal rolls are stable until one eigenvalue 
crosses the zero axis and the picture switches to the now emerging 
intermediate squares. This bifurcation and also that between intermediate 
and thermal squares becomes again imperfect for $\delta_i \neq 0$. 

For $\delta_i = 0$, at the beginning of the intermediate square interval two 
eigenvalues approach each other on a very small $\mu$--interval such that 
the curves look discontinuous here. When they meet they form a complex pair 
and only the real part is shown now. This complex pair becomes critical
in the middle of the interval and gives rise to the emergence of the 
oscillatory crossrolls. After the pair separates rapidly again one 
eigenvalues goes again through zero. The squares remain unstable because the 
other eigenvalue remains positive. 

Since for $\delta_i \neq 0$ the bifurcation point between solutal and 
intermediate squares vanishes, the first crossroll pair that emerges from 
the squares, the $A_1B_1A_2$-- and $A_1B_1B_2$--crossrolls must become 
disconnected from the square branch. But the oscillatory bifurcation and 
the other stationary bifurcation remain. For other parameters the scenario
is simpler. All eigenvalues stay real and only one crosses the zero axis.
When this is the case, only stationary crossrolls appear.

Note that the pair of eigenvalues responsible for the appearance of oscillatory 
crossrolls and the eigenvalue of the stationary crossrolls are not independent 
of each other. The latter emerges from the former. This is a feature that can 
also be observed in the full system of hydrodynamic field equations.

\begin{figure}
\leavevmode
\centering
\includegraphics[width=8cm]{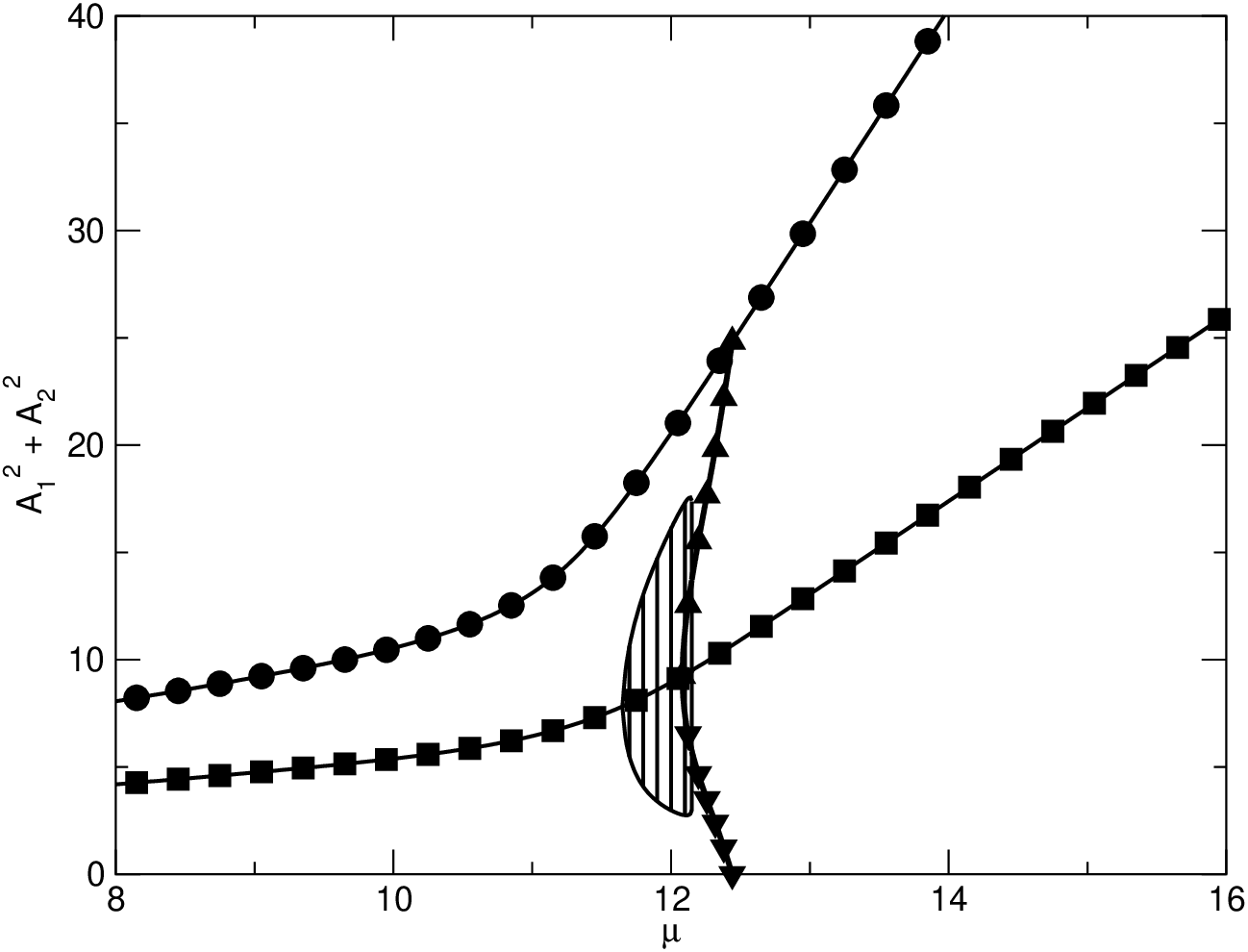}
\caption{Bifurcation diagram for $\delta_i \neq 0$. Symbols identify rolls,
squares, and stationary crossrolls as in Fig.~\ref{RBbif}. Vertical lines 
refer to oscillatory crossrolls. The stability properties are not displayed 
here.}
\label{amplibifdelta}
\end{figure}

A bifurcation diagram for slightly different parameters and especially 
larger symmetry--breaking terms $\delta_i$ is shown in 
Fig.~\ref{amplibifdelta}. The stability properties are the same as in 
Fig.~\ref{RBbif}. They are not displayed here.
The bifurcation diagram shows all the properties already deducted from the 
eigenvalues. The lower solutal, intermediate, and upper thermal branches have
fused to a smooth curve that nevertheless still shows the characteristic
upturning at the transition from the Soret to the Rayleigh regime. The 
$A_1B_1A_2$-- and $A_1B_1B_2$--crossrolls have become disconnected from the 
square branch. The other stationary crossroll bifurcation does still exist
and is preceded by an oscillatory crossroll bifurcation. Qualitatively, the
picture is exactly as in Fig.~\ref{RBbif}.

\begin{figure}
\leavevmode
\centering
\includegraphics[width=8cm]{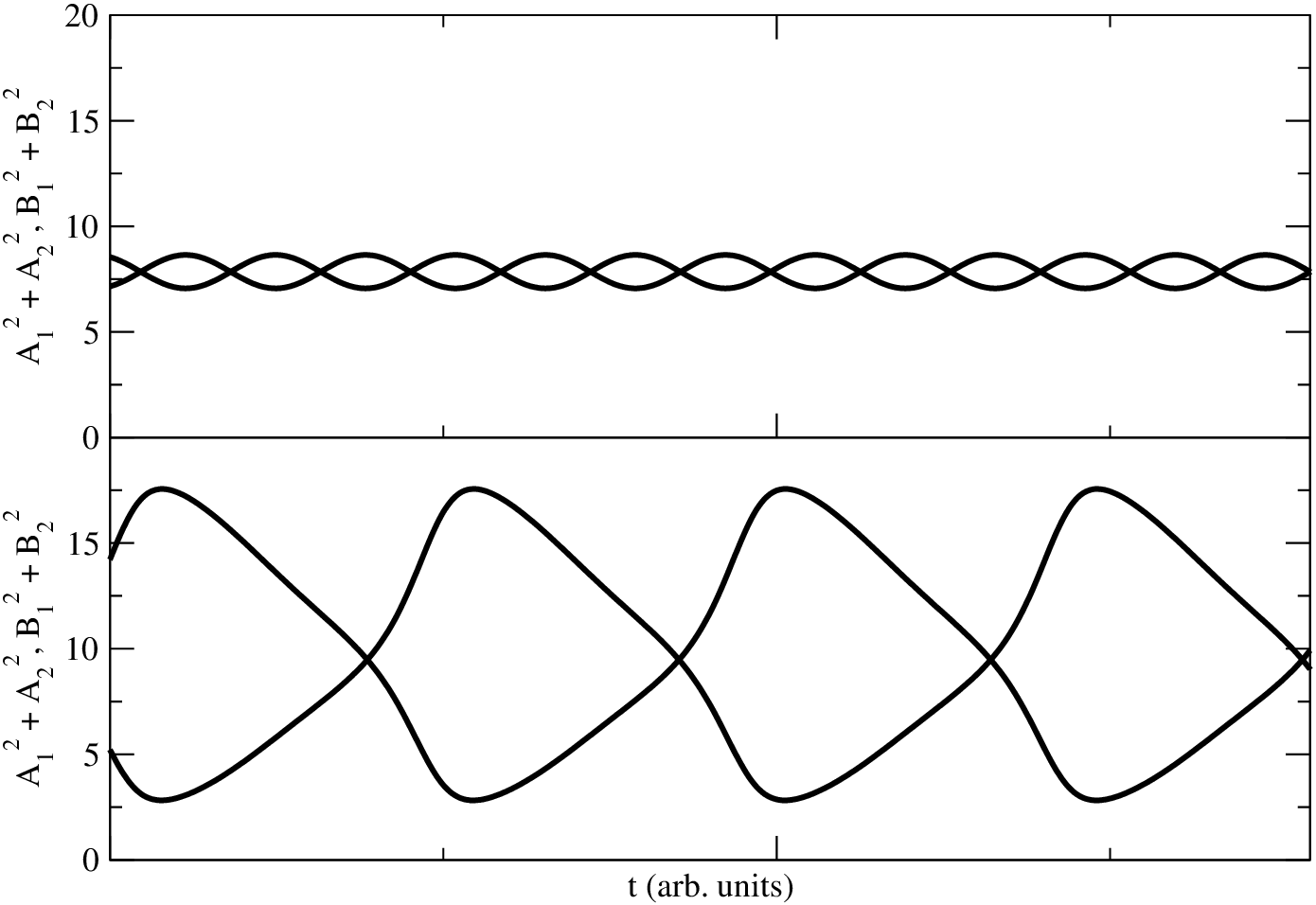}
\caption{Oscillatory crossrolls in the model for the parameters of
Fig.~\ref{amplibifdelta} and $\mu = 11.66$ (top), $\mu = 12.14$ (bottom).}
\label{oscimodel}
\end{figure}

Before we conclude, we want to take a closer look at the dynamics of the 
oscillatory 
crossrolls. Fig.~\ref{oscimodel} shows oscillations near the beginning 
(upper plot) and the end (lower plot) of the oscillatory crossrolls branch. 
Comparing to Fig.~\ref{oscifull} one sees the same behavior: An 
oscillation in counterphase with growing amplitude, decreasing frequency, and 
growing anharmonicity.

The details of the entrainment process that lead from oscillatory crossrolls
to the stationary type shall not be discussed here, but they are 
different from what we found in the 
numerical simulations of the full system of field equations \cite{JHL98} although 
similar transitions have been found in experiments  \cite{BCC90}. 

We found that
the details of the dynamic behavior of the model system (\ref{fulleqns2})
depends significantly on the chosen parameters. Thus, at present we conclude 
that our model equations (\ref{fulleqns2})
are capable of qualitatively reproducing the bifurcation scenario of 
Fig.~\ref{RBbif} and the dynamics of crossrolls in Fig.~\ref{oscifull} as
demonstrated in Figs.~\ref{amplibifdelta} and \ref{oscimodel}.

\section{Conclusion}

We have investigated the Rayleigh--B{\'e}nard convection in binary mixtures
with a positive separation ratio. We have shown that the observed bifurcation
scenario at small $L$, involving square patterns, rolls, stationary and 
oscillatory crossrolls arises naturally out of a simple system of two
coupled amplitude equation systems both consisting of two cubic equations
describing the dynamic of rolls in $x$-- and $y$--direction respectively. 
Attempts to extract such a model from the basic equations are underway.

~

{\it Acknowledgement} This work was supported by the Deutsche 
Forschungsgemeinschaft. We dedicate it to Prof. Dr. Siegfried Gro{\ss}mann on the
occasion of his 75th birthday.

\end{document}